\documentclass[12pt]{article}
\usepackage{amssymb,epsfig}
\usepackage{amsmath}

\newcommand{\be}{\begin{equation}}
\newcommand{\ee}{\end{equation}}
\newcommand{\bea}{\begin{eqnarray}}
\newcommand{\eea}{\end{eqnarray}}
\newcommand{\ba}{\begin{array}}
\newcommand{\p}[1]{(\ref{#1})}
\newcommand{\ea}{\end{array}}

\def\bbox{{\,\lower0.9pt\vbox{\hrule \hbox{\vrule height 0.2 cm
\hskip 0.2 cm \vrule height 0.2 cm}\hrule}\,}}
\newcommand{\dsl}{\pa \kern-0.5em /}

\newcommand{\dd}{{\mathrm d}}
\newcommand{\e}{{\mathrm e}}
\newcommand{\de}{\partial}
\newcommand{\refeq}[1]{(\ref{#1})}
\newcommand{\vv}{\wedge}

\newcommand{\bep}{\mbox{\boldmath $\epsilon$}}

\newcommand{\BI}{\mathrm{I}}
\newcommand{\BK}{\mathrm{K}}
\newcommand{\nn}{\nonumber \\}
\newcommand{\bbR}{\mathbb{R}}

\newcommand{\trsp}{\mathrm{T}}
\newcommand{\rep}[1]{\mathbf{#1}}
\newcommand{\CY}{CY}
\newcommand{\Ka}{K\"{a}hler}
\newcommand{\HK}{hyper--K\"{a}hler}
\newcommand{\spin}{{\slshape Spin(7)}}
\newcommand{\gman}{$G_2$}

\begin{document}
\begin{titlepage}

\vfill
\begin{flushright}
QMUL-PH-01-10\\
hep-th/0110034\\
\end{flushright}
\begin{center}
\vskip 1cm
{\bf \Large Fivebranes Wrapped on SLAG Three-Cycles}
\vskip 5mm
{\bf \Large and Related Geometry}
\vskip 10.mm
{Jerome P. Gauntlett$^{*1}$, Nakwoo Kim$^{\dagger 2}$, Dario Martelli$^{*3}$
and Daniel Waldram$^{*4}$}
\vskip 0.5cm
$^*$ {\it Department of Physics\\
Queen Mary \& Westfield College\\
University of London\\
Mile End Rd, London E1 4NS, UK}\\
\vskip 0.5cm
$^\dagger${\it Max-Planck-Institut f\"ur Gravitationsphysik\\
Albert-Einstein-Institut\\
Am M\"uhlenberg 1, D--14476 Golm, Germany}
\\
\vspace{6pt}
\end{center}
\par
\begin{abstract}
\noindent
We construct ten-dimensional supergravity solutions corresponding to the near
horizon limit of IIB fivebranes wrapping special Lagrangian three-cycles of
constant curvature. The case of branes wrapping a three-sphere provides
a gravity dual of pure ${\cal N}=2$ super-Yang-Mills theory in $D=3$.
The non-trivial part of the solutions are seven manifolds
that admit two $G_2$ structures each of which is covariantly
constant with respect to a different connection
with torsion. We derive a formula for the generalised
calibration for this general class of solutions.
We discuss analogous aspects of the geometry that arises when fivebranes
wrap other supersymmetric cycles which lead to $Spin(7)$ and 
$SU(N)$ structures.
In some cases there are two covariantly constant structures and in others
one. 
\end{abstract}

\vfill
\hrule width 5.cm
\vskip 5mm
{\small
\noindent $^1$ E-mail: j.p.gauntlett@qmw.ac.uk \\
\noindent $^2$ E-mail: kim@aei-potsdam.mpg.de\\
\noindent $^3$ E-mail: d.martelli@qmw.ac.uk \\
\noindent $^4$ E-mail: d.j.waldram@qmw.ac.uk
}
\end{titlepage}

\section{Introduction}
\label{intro}

The near horizon limit of supergravity solutions corresponding
to branes wrapping supersymmetric cycles provide gravity duals
to the field theories arising on the branes. By exploiting
the fact that the field theories are ``twisted'' \cite{bvs},
the corresponding supergravity solutions can be constructed
first in an appropriate gauged supergravity and then uplifted
to $D=10$ or $D=11$. This was first demonstrated in \cite{malnun} and
has been generalised to a number of different cases in
\cite{malnuntwo} - \cite{Gauntlett:2001jj}.

An interesting class of examples to study is type IIB
NS- or D-fivebranes wrapping supersymmetric cycles, since
one can obtain supersymmetric Yang-Mills (SYM) theory on the unwrapped part
of the fivebrane in the IR. ${\cal N}=1$ SYM in $D=4$ was studied in \cite{malnuntwo},
${\cal N}=1$ SYM in $D=3$ in \cite{agk,Schvellinger:2001ib,Maldacena:2001pb} and
${\cal N}=2$
SYM in $D=4$ in \cite{gkmw,zaf}.
One of the purposes of this paper is to extend these investigations to
${\cal N}=2$ SYM in $D=3$. To do so we consider fivebranes wrapping special
Lagrangian (SLAG) three-cycles in Calabi--Yau threefolds. As discussed
in \cite{malnun} the appropriate limit to decouple gravity
implies that only the local geometry of the SLAG three-cycle is
important.

If we wrap the fivebranes on a SLAG three-sphere, for which 
the deformed conifold is the appropriate local model of the
geometry without the backreaction, we obtain
pure ${\cal N}=2$ SYM in $D=3$. Perturbatively this theory has a Coulomb branch
but a superpotential is generated by instanton (monopole) effects
\cite{ahw,deBoer,aha} and there is no stable vacuum. As such it might
seem impossible to obtain a gravity dual. However, these instanton effects
are suppressed in the large $N$ limit and might not be expected to survive
in the supergravity approximation. This was observed in the context of
${\cal N}=2$ SYM in $D=4$ in \cite{gkmw}, where the supergravity solutions
incorporated all perturbative effects but not the non-perturbative corrections.
Indeed we will construct singular supergravity solutions and by
probing these solutions with a fivebrane argue that they
describe a slice of the perturbative 
Coulomb branch of ${\cal N}=2$ SYM in $D=3$.

If a Chern-Simons term with suitable co-efficient is added to
pure ${\cal N}=2$ SYM in $D=3$, one expects that there is a unique
confining ground state \cite{ohta}. {}Following
\cite{malnuntwo,Schvellinger:2001ib,Maldacena:2001pb}
one then anticipates that a regular supergravity
solution dual to these theories should exist, which should
include non-zero NS flux on the SLAG three-sphere
to account for the Chern--Simons term~\cite{agk}.
The present work can be viewed as
a first step toward the construction of these solutions in the
same way that the work of \cite{agk} led to
\cite{Schvellinger:2001ib,Maldacena:2001pb} (who used \cite{chamvolk})
for the confining ${\cal N}=1$ $D=3$ theories. 

The non-trivial part of the $D=10$ solutions constructed here
are seven-dimensional manifolds with non-vanishing NS three-form $H$ and
dilaton. Defining two connections with totally anti-symmetric
torsion $\nabla^\pm =\nabla\pm\frac{1}{2}H$, where $\nabla$ is the Levi-Civita
connection, we show that the holonomy of each of these connections is
in $G_2$. In particular we show that the seven manifolds admit two $G_2$
structures, specified by associative three forms, one of which is covariantly constant
with respect to $\nabla^+$ and the other with respect to $\nabla^-$.
We also show that this is a general result for type IIB (and type IIA)
backgrounds preserving this amount of supersymmetry. In addition
we present the appropriate notion of generalised calibration. In particular
we derive an expression for the dual six-form potential (which has a
seven-form field strength that is dual to the three form $H$) in terms
of either of the $G_2$ structures and the dilaton.

A similar phenomenon was observed in \cite{gkmw} for fivebranes wrapping
a two-cycle in Calabi--Yau two-folds. {}For this case the non-trivial part
of the geometry was six-dimensional. It was shown that the manifold
admits two commuting complex structures and one of these is
covariantly constant with respect to $\nabla^+$ and the other
with respect to $\nabla^-$. Moreover, the dual six-form potential can be
constructed from either of these complex structures and the dilaton.
By contrast a different kind of six dimensional
geometry arises when fivebranes wrap a two-cycle inside a Calabi--Yau
threefold. It was shown in \cite{papatse} that the geometry found
in \cite{malnun} admits just one complex structure that is covariantly
constant with respect to one of the connections with torsion.
The dual six-form potential can still be
constructed from this complex structure and the dilaton.
The reason for the difference between the resulting two kinds of geometry
in these two cases is that in the former case there are overall transverse
directions (two) to the fivebrane wrapping the two-cycle, whereas in
the latter case, there are none. This means that after incorporating the
back-reaction of the fivebrane on the geometry, in the former case the
non-trivial manifold jumps from four dimensions, the Calabi--Yau two-fold,
to six as in the solution \cite{gkmw,zaf},
whereas in the latter example it starts at six, the Calabi--Yau
threefold, and remains at six, in the full solution \cite{malnun}.

We show that this holds more generally.
{}For example, when fivebranes wrap SLAG
three-cycles there is one  overall transverse direction and the
corresponding solutions are seven-dimensional with two $G_2$-structures.
By contrast when fivebranes wrap
associative three-cycles in manifolds
with $G_2$ holonomy there are no overall transverse directions and
the corresponding non-trivial seven manifold
\cite{agk,Schvellinger:2001ib,Maldacena:2001pb}
has a single $G_2$ structure.
We also discuss other cases that include manifolds with $SU(N)$ and
$Spin(7)$ structures and derive the expression for the
generalised calibrations.

\bigskip\noindent
{\bf Note:} In the process of writing up this paper, we became aware of
\cite{gr}. The results of that paper have some overlap with sections
\ref{generalities}, \ref{eqsmotion}, \ref{asymptotics} and
\ref{probecomput} of the work presented here.

\section{NS fivebranes wrapped on a SLAG three-sphere}
\label{generalities}

To obtain an $\mathcal{N}=2$ super-Yang--Mills theory in three
dimensions, we consider type IIB fivebranes wrapped on a
special Lagrangian three-cycle in a Calabi--Yau threefold. We start by
recalling that in the limit that the string coupling is set to zero,
keeping the string scale fixed, a configuration
of $N$ IIB NS fivebranes is
described by the six-dimensional IIB little string
theory~\cite{LST}. This theory flows in the IR to $D=6$
super-Yang--Mills theory. {}Further, it preserves sixteen supercharges
transforming as $({\bf 4}_+,{\bf 2_+})+({\bf 4}_-,{\bf 2_-})$ under
$SO(1,5)\times SO(4)_R$ where $SO(4)_R$ is the group of
$R$-symmetries. Geometrically this group describes rotations in
directions normal to the brane. In the large $N$ limit, the theory has
a gravity dual, given by the near horizon limit of $N$ NS-fivebranes,
namely,
\begin{equation}
\begin{aligned}
   \dd s^2 &= \dd\xi_{1,5}^2
       + N \left( \dd \rho^2 + \dd \Omega_3^2 \right) , \\
   \e^{-2\Phi} &= \e^{-2\Phi_0} \e^{2\rho} ,
\end{aligned}
\label{eq:flatNS5}
\end{equation}
where $\dd\Omega_3^2$ is the metric on a three-sphere, 
$\dd\xi_{1,5}^2$ is the Minkowski metric on $\bbR^{1,5}$ and
we have set $\alpha'=1$. There is
also a NS three-form flux through the three-sphere, normalised so that
the integral of the three-form $H/4\pi^2$ over the sphere is $N$. 

To obtain pure ${\cal N}=2$ Yang--Mills theory in $D=3$ in the IR
we consider this little string theory compactified on
$\bbR^{1,2}\times S^3$. The $SO(1,5)$ Lorentz group of the fivebrane
is then broken to $SO(1,2)\times SO(3)$. Naively, such a
compactification breaks all the supersymmetries. However, by
considering a twisted theory four supercharges can be preserved. To do
this, we split the $R$-symmetry $SO(4)_R\to SO(3)_R$ and then identify
the $SO(3)$ spin connection of the three-sphere with
$SO(3)_R$. Geometrically this is exactly the same twisting that arises
in the local description of a fivebrane wrapping a special Lagrangian
three-cycle. In this setting the $SO(3)$ part of the $R$-symmetry
corresponds to the symmetry group of the normal bundle to the
fivebrane inside the Calabi--Yau threefold.

It is straightforward to see that this twisting indeed preserves
${\cal N}=2$ supersymmetry in $D=3$. {}First note that the preserved
supersymmetries of the fivebrane transform as two copies of
$(\mathbf{2,2,2})$ under $SO(1,2)\times SO(3)\times SO(3)_R$. After
twisting these transform as two copies of $({\bf 2,1})+({\bf 2,3})$ 
under $SO(1,2)\times SO(3)_D$, where $SO(3)_D$ is the diagonal
subgroup of the two $SO(3)$ factors. It is then the singlets of
$SO(3)_D$ which are the preserved supersymmetries. The four scalars of
a single fivebrane transform as a ${\bf 4}$ of $SO(4)_R$ and hence as
a triplet and a singlet of $SO(3)_R$. Geometrically, this simply
corresponds to the split of the four-dimensional space transverse to
the branes into three directions within the Calabi--Yau threefold and one
remaining flat direction. The non-trivial twisting of the normal
directions means that the triplet of scalars really describe a section
of the normal bundle to the SLAG three-cycle within the Calabi--Yau
threefold. Given the identification of $SO(3)_R$ with the $SO(3)$
rotations in the tangent space of the cycle, they can also be viewed
as one-forms, that is, sections of the cotangent bundle of the
three-cycle. This matches the standard result that the normal
deformations of a SLAG three-cycle are given by harmonic one-forms on
the three-cycle~\cite{maclean}. As there are no harmonic one-forms on
a three-sphere, it is necessarily a rigid SLAG three-cycle within a
Calabi--Yau threefold and the four scalar fields give rise to one
real massless scalar in $D=3$ coming from the singlet. The
gauge fields on the fivebrane also have no scalar zero-modes on the
three-sphere and thus simply give rise to a $D=3$ gauge field. 
These fields plus the fermionic partners comprise the field content
of pure $D=3$ ${\cal N}=2$ $U(1)$ Yang-Mills theory. Generalising to
$N$ fivebranes gives rise to $SU(N)$ gauge group. 

Note that we could also consider wrapping on other constant
curvature cycles. In particular, we would get the same theory in the
IR if we considered fivebranes wrapping a lens space,
$S^3/\Gamma$. Alternatively, if we wrapped on a torus $T^3$, we would
get three additional chiral adjoint matter multiplets from zero 
modes of the scalar triplet and scalar zero modes of the gauge
fields. In this case, there are actually 16 preserved supercharges,
the matter and gauge fields form a single multiplet and the gauge theory
in the IR is simply the dimensional reduction of $D=6$
super-Yang--Mills theory. {}Finally, we could also consider the
fivebranes wrapping a compact space constructed as a quotient of
hyperbolic space, $H^3/\Gamma$. In this case, we again get
additional $\mathcal{N}=2$ chiral matter multiplets in the adjoint 
representation. We will briefly mention the structure of the
supergravity solutions for $H^3$ in the next section, but will
otherwise concentrate on the $S^3$ case.

\section{Supergravity solution}
\label{eqsmotion}

We first note that the IIB supergravity solutions describing wrapped
NS fivebranes have only NS-NS fields non-vanishing and hence are also
solutions of ${\cal N}=1$ supergravity in $D=10$. To explicitly
construct the near horizon limits, we follow the strategy
of~\cite{malnun} and first construct the solutions within a suitable
gauged supergravity and then uplift to $D=10$. The relevant $D=7$
gauged supergravity is obtained from the consistent truncation of
${\cal N}=1$ supergravity in $D=10$ on a three-sphere. This gives the
$D=7$ $SO(4)$ gauged supergravity of \cite{salsez} with the gauge fields
arising from the isometries of the three-sphere. A reader not
interested in the methods used to derive the solution, could skip the
first subsection and begin directly with the full solution in ten
dimensions. We also note that given this solution the corresponding
solution for wrapped D-fivebranes can then be simply obtained by
S-duality.

\subsection{Supergravity solution in D=7 gauged supergravity}
\label{HJ}

Starting from the bosonic Lagrangian of the $D=7$ $SO(4)$ gauged
supergravity, let us derive BPS equations for the wrapped branes
using the effective Lagrangian method given in~\cite{gkmw}.
One can also check that the same BPS equations are obtained directly
from the supersymmetry variations as given in~\cite{salsez}, and that
solutions of the BPS equations preserve 1/8 of the supersymmetry.
The bosonic field content of the gauged supergravity~\cite{salsez}
(see also~\cite{clp}) comprises of a metric, gauge fields $F^{ab}$ in
the adjoint of $SO(4)$, a three-form and ten scalar fields in a
symmetric four-by-four matrix $T_{ij}$. 

Geometrically the gauge fields describe the twisting of the normal
bundle to the fivebrane. Thus to incorporate the twisting discussed in
the last section we need to have non-vanishing gauge fields in
an $SO(3)$ subgroup of $SO(4)$. We realise this by defining $F^a\equiv
\frac{1}{2}\epsilon^{abc}F^{bc}$ for $a,b,c=1,2,3$ and set $F^{a4}=0$. An
ansatz for the scalar fields consistent with the twisting, preserving
$SO(3)\subset SO(4)$ is given by
\bea
T_{ij} & = &\e^{y/4}\mathrm{diag}(\e^x,\e^x,\e^x,\e^{-3x})~. 
\eea
{}Finally, we set the seven dimensional three-form to zero. It is
important to check that such an ansatz is consistent with the
three-form, scalar and metric equations of motion, as is easily
verified in this case. 

With this ansatz the equations of motion
are encoded in the following seven dimensional Lagrangian  
\bea
{\cal L} =  \sqrt{g}\left\{R -\frac{5}{16}\de_\mu y \de^\mu y -
3\de_\mu x \de^\mu x -\frac{1}{4} \e^{-y/2-2x}
F^a_{\mu\nu} {}F^{a\,\mu\nu} \right. & & \nonumber\\
\left. + \frac{1}{2}g^2
\e^{y/2}(3\e^{2x}-\e^{-6x}+6\e^{-2x})\right\}& &
\label{7dlagrangian}
\eea
which can be derived from~\cite{salsez} or~\cite{clp}. {}For the metric
and gauge fields we use the ansatz
\bea
\dd s^2 & = & \e^{2f(r)}(\dd\xi^2 + \dd r^2) +
\frac{a^2(r)}{4}(\sigma_1^2+\sigma_2^2+\sigma_3^2) \nonumber\\
A^{a} & = & \frac{1}{2g}\sigma^a\label{twisting}
\eea
where $\dd\xi^2$ is the Minkowski metric on $\bbR^{1,2}$, and
the left-invariant one-forms of $SU(2)$ satisfying
$\dd \sigma^a = \frac{1}{2}\epsilon^{abc}\sigma^b\vv\sigma^c$ give the
round metric on $S^3$. In particular note that we have explicitly
incorporated the required twisting by setting the $SO(3)$ gauge
fields equal to the $SO(3)$ spin connection of the three-sphere. 

When we uplift to $D=10$, in order to find solutions corresponding to
wrapped NS-fivebranes, we require that the string-frame warp factor of
the $\dd\xi^2$ term is unity. This implies that, as in \cite{gkmw}, we
set $y=-4f$. Substituting this ansatz into the
Lagrangian~\refeq{7dlagrangian} we obtain a one-dimensional effective
action which in terms of the new variables
\bea
\e^{2A}  =  \e^{2f} a^3~, \qquad\qquad \e^{2h}  =  \e^{-2f} a^2~,
\eea
is written as
\bea
L & = & \e^{2A}\left[ 4\dot A^2 - 3 \dot h^2 - 3\dot x^2 -V \right]\nn
V(h,x) & = & -6\e^{-2h} + \frac{3}{2g^2}\e^{-4h-2x} -
\frac{g^2}{2}(3\e^{2x}+6\e^{-2x}-\e^{-6x})~.
\label{effectivelag}
\eea
The equations of motion derived from this Lagrangian, combined with
the constraint of setting the Hamiltonian
to zero (arising from diffeomorphisms in the radial variable) give
the differential equations satisfied by $A$, $h$, and $x$.
This system can then be reduced to a set of first-order Hamiltonian
equations together with an associated Hamilton--Jacobi
equation. Choosing ${}F=\e^{2A}W(h,x)$ as a principle function, the
Hamilton--Jacobi equation gives rise \cite{gkmw} to the following
non-linear equation for the superpotential $W$ 
\bea
V & = & \frac{1}{4}\left(\frac{1}{3}\de_h W^2 +\frac{1}{3}\de_x W^2
 -W^2\right)~.
\label{superp:eq}
\eea
By inspection, one finds the solution
$W =  - g(\e^{-3x}+3\e^{x}+3g^{-2}\e^{-2h-x})$, from which
the first-order BPS equations then follow, and read
\bea
\dot A & = & \frac{1}{4}W=-\frac{g}{4}(\e^{-3x}+3\e^{x}+\frac{3}{g^2}\e^{-2h-x})\nonumber\\
\dot h & = & -\frac{1}{6}\de_h W = -\frac{1}{g}\e^{-2h-x}\\
\dot x & = & -\frac{1}{6} \de_x W=-\frac{g}{2}(\e^{-3x}-\e^{x}+\frac{1}{g^2}\e^{-2h-x})~.
\label{BPS}
\eea
These equations can be solved explicitly by introducing a new
radial variable given by $z = \frac{1}{2}g^2\e^{2h}$. The solution is
\bea
\e^{-2x} & = & \frac{\BI_\frac{3}{4}[z]-c
\BK_\frac{3}{4}[z]}{\BI_{-\frac{1}{4}}[z]+c
\BK_\frac{1}{4}[z]}\label{inve2x}\\
\e^{A+\frac{3}{2}x} & = & z \left(\BI_{-\frac{1}{4}}[z]+c
\BK_{\frac{1}{4}}[z] \right)~,
\eea
where K$_\nu$ and I$_\nu$ are the modified Bessel functions and $c$
is an integration constant. A second integration constant
appears in the second equation, but it can be set to unity
by a coordinate transformation in the metric.

\begin{figure}[!th]
\vspace{5mm}
\begin{center}
\epsfig{file=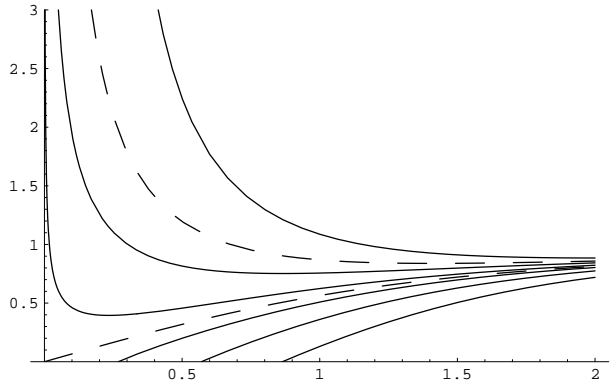,width=8cm,height=5cm}\\
\end{center}
\caption{Plot of $\e^{-2x}$ versus the radial coordinate $z$, for the
three-sphere case. The lower
dashed line corresponds to $c=0$, whereas the upper one corresponds to
$c=-\sqrt{2}/\pi$.}
\label{fig1}
\vspace{5mm}
\end{figure}

We have plotted the various orbits of $\e^{-2x}$ in {}Figure \ref{fig1}
labelled by different values of $c$. We will see in the next section
that these correspond to the flows from the UV to the IR at different
points in the moduli space of the gauge theory. {}For the moment, we
simply note that there are three distinct regions. {}For $c\ge0$ there is
a singularity in the solution at finite $z=z_0\ge0$ where $\e^{-2x}$
vanishes. {}For $-\sqrt{2}/\pi\leq c< 0$ the singularity
occurs at $z=z_0=0$, and now $\e^{-2x}$ diverges. {}For $c<-\sqrt{2}/\pi$
there is again a singularity where $\e^{-2x}$ diverges but now at a
finite non-zero value of $z$. 

\begin{figure}[!th]
\vspace{5mm}
\begin{center}
\epsfig{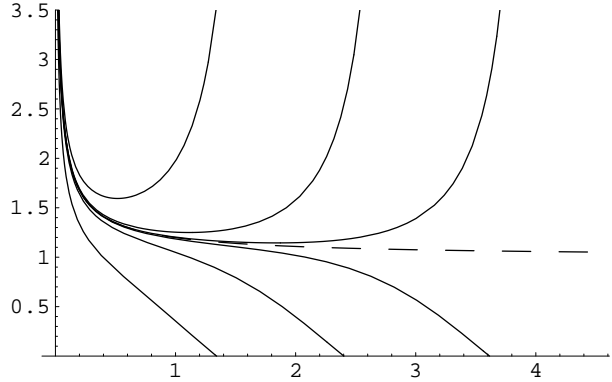}\\
\end{center}
\caption{Plot of $\e^{-2x}$ versus the radial coordinate $z$, for the
hyperbolic case. The dashed line corresponds to $c\to \infty$.}
\label{fig2}
\vspace{5mm}
\end{figure}

We can easily obtain solutions replacing the three-sphere with the
hyperbolic space $H^3$ or quotients thereof. The only change occurring
in the Lagrangian \refeq{effectivelag} is the sign in front of the
first term of the potential, induced by the negative curvature. The
solution for first order equations is obtained by formally replacing
$g^2\to -g^2$, and results in an overall sign change in \refeq{inve2x}.
This means that the hyperbolic space case corresponds to the negative branch
of the very same solutions depicted in {}Figure \ref{fig1} and we have plotted
them in Figure \ref{fig2}. As the dual field-theory 
interpretation of these solutions is not clear we will not discuss them
further in the following and instead focus on the three-sphere case.

\subsection{The supergravity solution in D=10}
\label{uplift}

Using the formulae given in \cite{clp} we can uplift the
seven-dimensional solution to obtain the corresponding $D=10$
supergravity solution. This can be viewed as an ${\cal N}=1$
solution or a type IIA or IIB solution with only NS-NS fields
non-vanishing.  

In the string frame our family of solutions read
\bea
\dd s^2 &=& \dd \xi^2 +
\frac{2z}{g^2}\dd \Omega_3^2 + \frac{\e^{2x}}{g^2}(\dd z^2 + \dd \psi^2) +
\frac{1}{g^2\Omega}\sin^2\psi (E_1^2 + E_2^2)
\nonumber\\
\eea
where
\bea
E_1 & = & \dd \theta + \cos\phi\frac{\sigma^1}{2}-
\sin\phi\frac{\sigma^2}{2}\nn
E_2 & = & \sin\theta \Big(\dd\phi +\frac{\sigma^3}{2}\Big)- \cos\theta
\Big(\sin\phi\frac{\sigma^1}{2}+\cos\phi\frac{\sigma^2}{2}\Big)
\eea
and
\bea
\Omega= \e^{2x} \sin^2\psi + \e^{-2x} \cos^2 \psi~.
\eea
As before, the coordinates $\{\xi^i\}$, $i=0,1,2$, parameterise the
Minkowski space on the unwrapped world volume directions of the
fivebrane while $\dd \Omega_3^2$ is the metric on the three-sphere on
which the fivebrane is wrapped and is expressed in terms of the
$SU(2)$ left-invariant one-forms $\sigma^a$. The coordinates
$0\le\psi\le\pi,0\le\theta\le\pi,0\le\phi<2\pi$ parameterize the
squashed and twisted three-sphere transverse to the fivebrane.

The dilaton is given by
\bea
\e^{-2\Phi+2\Phi_0} = \Omega\sqrt z\left(\BI_\frac{3}{4}[z]-c
\BK_\frac{3}{4}[z] \right)^2~,
\eea
and the NS three-form has the form
\bea
H  &=& -\frac{2\sin^2\psi}{g^2\Omega^2}\Big[(2\cos^2\psi\sinh^2 2x -
\e^{2x}\cosh 2x)\dd\psi\nn
&&\qquad\qquad\qquad 
  +\cos\psi\sin\psi (2\sinh 2x -\frac{1}{2z})\dd z\Big] E_1  E_2 \nn
&&- \frac{\sin\psi\cos\psi\e^{-2x}}{4g^2\Omega}\Big[
-\sin\theta\sigma^1\sigma^2 E^1
+\sigma^3\sigma^1 (\cos\theta\cos\phi E^1-\sin\phi E^2)\nn
&&\qquad\qquad\qquad\qquad
 + \sigma^2\sigma^3 (\cos\theta\sin\phi E^1+\cos\phi E^2)\Big]
\nn
&&- \frac{1}{4g^2}\Big[\cos\theta\sigma^1\sigma^2
+\sin\theta\cos\phi\sigma^3\sigma^1
+\sin\theta\sin\phi\sigma^2\sigma^3\Big] \dd\psi
\eea
where the wedge product of forms is understood. Note that the solution
depends on two parameters, the expectation value of the dilaton,
$\Phi_0$ and the integration constant $c$ appearing in $\e^{-2x}$
as given in \refeq{inve2x}.

\section{Supersymmetry and $G_2$ structure}
\label{susy}

We have checked that our solution preserves two of the 16 supercharges
in the $SO(4)$ gauged supergravity in $D=7$ and consequently 
also preserves two supercharges in type I supergravity in $D=10$, and
four supercharges as a type IIA or IIB solution. It is illuminating to
see how this works in $D=10$ as this will allow us to elucidate an
interesting notion of generalised calibration. We will concentrate on
the case of type IIB, though the corresponding type I and type IIA
solutions admit a similar analysis. 

We first introduce a rather non-obvious orthonormal frame given by
\begin{equation}
\begin{aligned}
   e^a &= \frac{\sqrt{2z}}{g} S^a~, \qquad a=1,2,3 & 
   e^7 &= \frac{1}{g\Omega^{1/2}}\left( \cos\psi \dd z
       -\e^{2x}\sin\psi\dd\psi\right) \\
   e^4 &= \frac{1}{g\Omega^{1/2}}\left( \sin\psi
       \e^{2x}\dd z+\cos\psi\dd\psi\right) \qquad & 
   e^8 &= \dd \xi^1 \\
   e^5 &= \frac{1}{g\Omega^{1/2}}\sin\psi E_1 & 
   e^9 &= \dd \xi^2 \\
   e^6 &= \frac{1}{g\Omega^{1/2}}\sin\psi E_2 & 
   e^0 &= \dd \xi^0 
\end{aligned} 
\label{eq:frame}
\end{equation}
where the one-forms $S^a$ are defined as
\bea
S^1 & = & \cos\phi \frac{\sigma^1}{2} -\sin\phi \frac{\sigma^2}{2}\nn
S^2 & = & \sin\theta \frac{\sigma^3}{2} -\cos\theta\Big( \sin\phi
\frac{\sigma^1}{2} +\cos\phi \frac{\sigma^2}{2}\Big)\nn
S^3 & = & -\cos\theta \frac{\sigma^3}{2} -\sin\theta\Big( \sin\phi
\frac{\sigma^1}{2} +\cos\phi \frac{\sigma^2}{2}\Big)~.
\eea
Note that $e^4$ and $e^7$ of this frame incorporate a rotation of
$\dd z$ and $\dd\psi$. Geometrically they describe two radial
directions, one in the space transverse to the brane within the
Calabi--Yau threefold and one in the remaining overall transverse
direction. A similar frame was used in \cite{gkpw,gkmw}. In addition
we note that $S^1$, $S^2$ and $S^3$ are related to $\sigma^1$,
$\sigma^2$ and $\sigma^3$ by an $SO(3)$ rotation parametrised by the
coordinates $\theta,\phi$ of the transverse three-sphere. In this
frame the NS three-form is given by 
\bea\label{eq:Hframe}
H&=&
\frac{g\e^{-2x}}{2z\Omega^{1/2}}\Big[\cos\psi(e^{124}-e^{236}-e^{135})
-\e^{2x}\sin\psi e^{127}\Big]\nn
&&-\frac{g\e^{-2x}\sin\psi}{\Omega^{3/2}}\Big[\e^{6x}\sin^2\psi
+\e^{2x}(4\cos^2\psi+1)
-3 \e^{-2x}\cos^2\psi-\frac{1}{z}\cos^2\psi\Big] e^{567}\nn
&&-\frac{g\e^{-2x}\cos\psi}{\Omega^{3/2}}\Big[\e^{4x}\sin^2\psi-3+
\e^{-4x}\cos^2\psi-\frac{\e^{2x}}{z}\sin^2\psi\Big]e^{456}~,
\eea
where $e^{mnp}=e^m \wedge e^n \wedge e^p$. 

The type IIB supersymmetry transformations are given by
\bea
\delta \lambda &=&  \Gamma^\mu\de_\mu \Phi \tau_3 \bep -
\frac{1}{12}H_{\mu\nu\rho}
\Gamma^{\mu\nu\rho}\bep = 0 \nonumber\\
\delta \psi_\mu &=& \nabla_\mu \bep
-\frac{1}{8}H_{\mu\nu\rho}\Gamma^{\nu\rho}\tau_3\bep = 0~,
\label{gravitino}
\eea
where $\bep=(\epsilon^-,\epsilon^+)$ is the $SO(2)$-doublet of chiral
IIB supersymmetry parameters and $\tau_3$ is the third Pauli
matrix. Note that the gravitino variation can be written in component
form as 
\begin{equation}
   \nabla^-\epsilon^- = 0~, \qquad \nabla^+\epsilon^+ = 0~,
\end{equation}
where we introduce a generalized connection with totally antisymmetric
torsion, given by
\begin{equation}
   \label{eq:gencon}
   \nabla^{\pm}_\mu = \nabla_\mu 
        \pm \frac{1}{8}H_{\mu\nu\rho}\Gamma^{\nu\rho}~,
\end{equation}
where $\nabla_\mu$ is the Levi--Civita connection.
{}From the dilatino variation we infer the following projections on the
preserved supersymmetry 
\bea\label{projection}
\Gamma^{1256} \bep & = &\bep \nn
\Gamma^{1346} \bep & = & \bep \nn
\Gamma^{4567} \bep & = & \tau_3\bep~.
\eea
We will not give the angular dependence of the spinors which
can be determined by examining the gravitino variation.
These projections preserve four independent Killing spinors,
corresponding to ${\cal N}=2$ in $D=3$. 
Note that the first two projections are the same as
those for Killing spinors on a Calabi--Yau threefold with tangent
directions $\{1,2,3,4,5,6\}$ and the third projection for a IIB NS
fivebrane with tangent directions $\{0,1,2,3,8,9\}$. In other words,
the supersymmetry preserved matches that of a NS fivebrane probe wrapping
a SLAG three-cycle in a Calabi--Yau threefold as expected. Note that
the analogous analysis shows that as a solution of type IIA it also
preserves four supercharges while as a solution of type I it preserves
just two. 

The projections \refeq{projection} 
correspond to those of a $G_2$ holonomy
manifold with tangent directions $\{1,2,3,4,5,6,7\}$. The presence of
$\tau_3$ means that the projections for $\epsilon^-$ and $\epsilon^+$
differ by a sign. The two are related by simply reversing the sign of
$e^7$ in the orthonormal frame~\eqref{eq:frame}. 
This structure gives us an alternative interpretation of the
solution, which is discussed in more generality in
section~\ref{gencal}. The seven-dimensional part of the metric can be 
viewed a $G_2$ holonomy manifold with totally antisymmetric
torsion. In fact, the manifold admits two distinct covariantly
constant spinors $\epsilon^-$ and $\epsilon^+$ with respect to
two distinct connections with totally antisymmetric torsion, 
$\nabla^\pm$, each of which has $G_2$ holonomy. 
The presence of $G_2$ holonomy can be characterised by the existence of a
covariantly constant associative three-form $\phi$. This can be
constructed from the covariantly constant spinor $\epsilon$ by 
\begin{equation}
   \label{eq:phi}
   \phi_{\alpha\beta\gamma} =
       \epsilon^\mathrm{T}\gamma_{\alpha\beta\gamma}\epsilon~,
\end{equation}
where $\gamma_\alpha$ are gamma matrices on the $G_2$ holonomy
manifold, and we normalize $\epsilon$ by $\epsilon^\trsp\epsilon=1$. 
Thus, since we have two different spinors $\epsilon^\pm$,
we can construct two different associative three-forms. Given the
projections~\eqref{projection}, these can be immediately written down in
the frame~\eqref{eq:frame}, giving 
\bea\label{eq:phii}
\phi^- &=&  e^{123}+ e^{145} - e^{167}
+e^{246} + e^{257} +e^{347} - e^{356}\nn
\phi^+ &=&  -e^{123}- e^{145} - e^{167}
-e^{246} + e^{257} +e^{347} +e^{356}~.
\eea
Since $\nabla^-\epsilon^-=0$ and $\nabla^+\epsilon^+=0$, then, by
construction, the associated three-forms are covariant constant but
with respect to different connections
\bea
\nabla^-\phi^- &=& 0\nn
\nabla^+\phi^+ &=& 0~.
\eea

We should also be able to view the associative three-forms $\phi^\pm$ as
a generalized calibration. Again this will be discussed in more detail
in section~\ref{gencal}. {}For the moment we simply note that using the
explicit expressions~\eqref{eq:Hframe} and~\eqref{eq:phii}, one can
derive an expression for the dual NS six-form potential $\tilde{B}$
defined by $\dd\tilde{B}\equiv\tilde{H}=*\e^{-2\Phi}H$. We find that 
\bea
\tilde{B} & = & \pm \mathrm{Vol_3} \vv \e^{-2\Phi} \phi^\pm~,
\label{NSsixform}
\eea
where $\mathrm{Vol_3}$ is the volume form of the unwrapped part
of the fivebrane world-volume. This
holds for \textit{either} choice of $\phi^\pm$, the two expressions
differing by a gauge transformation. This expression for the dual
potential will be particularly useful when we come to consider a probe
calculation in section~\ref{probecomput}. It is equivalent to  
\begin{equation}
   \label{eq:gencalG2}
   \dd \left( \mathrm{Vol_3} \wedge \e^{-2\Phi} \phi^\pm \right) 
       = \pm\tilde{H}~,
\end{equation}
which, as we will show, can be viewed as the generalized calibration
condition on a $G_2$ holonomy manifold with torsion and a non-trivial
dilation field.

\section{UV and IR limits}
\label{asymptotics}

Let us now discuss the asymptotic UV and IR behaviour of our solutions.
The UV limit is obtained when $z\to \infty$. The metric and dilaton
then take the form
\bea
\dd s^2 &\approx& \dd \xi^2 +
\frac{2z}{g^2}\dd \Omega_3^2 + \frac{1}{g^2}[\dd z^2 + \dd \psi^2
+\sin^2\psi (E_1^2 + E_2^2)]\nonumber\\
\e^{-2\Phi+2\Phi_0} &\approx& \frac{1}{2\pi \sqrt{z}} \e^{2z}~,
\eea
This has the same form as the near horizon limit of the flat
NS-fivebrane solution~\eqref{eq:flatNS5} but with world volume
$\bbR^{1,2}\times S^3$ instead of $\bbR^{1,5}$ and the appropriate twisting.
Note that the dilaton is asymptotically linear up to a logarithmic
correction. 

\begin{figure}[!th]
\vspace{5mm}
\begin{center}
\epsfig{file=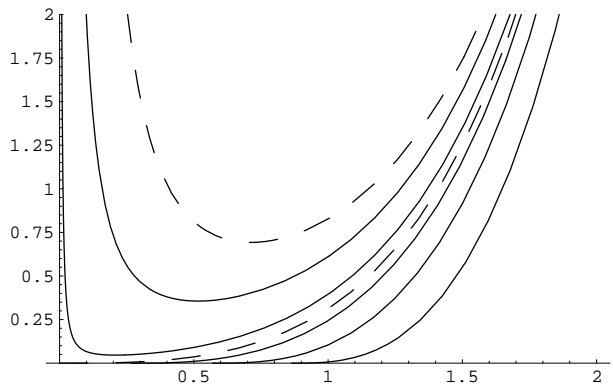,width=8cm,height=5cm}\\
\end{center}
\caption{Plot of $\e^{-2\Phi}$ versus the radial coordinate $z$. The lower
dashed line corresponds to $c=0$, whereas the upper one corresponds to
$c=-\sqrt{2}/\pi$.}
\label{fig3}
\vspace{5mm}
\end{figure}

The one parameter family of solutions, specified by $c$, are all
singular in the IR, that is for $z\approx z_0$, where $z_0$
is defined where $\e^{-2x}$ goes to zero or diverges, depending
on the value of $c$ (see {}Figure 1). This can be seen by analysing 
the behaviour of the dilaton, which is plotted in {}Figure \ref{fig3}
(for $\psi=0$). 
{}{}For the solutions with $c<0$, it blows up at $z_0$ and
generic values of $\psi$ while for the
solutions with $c\geq 0$ it vanishes at $z_0$ for generic $\psi$.
Since the $g_{tt}$ component of the Einstein frame metric is given by
$\e^{-\Phi/2}$, these singularities are of the ``good'' type for  $c\geq0$
and of the ``bad'' type for $c< 0$, using the criteria of \cite{malnun}.
We thus expect that only the solutions with $c\geq0$ are associated with
${\cal N}=2$ $D=3$ Yang-Mills theory phenomena. 

We now analyse the limiting form of the metric near the good
singularities.
{}First notice that the function $\BI_{-\frac{1}{4}}+c\BK_{\frac{1}{4}}$ is
positive definite for $c\geq 0$, and hence a zero of $\e^{-2x}$ occurs when
\bea
\BI_\frac{3}{4}[z_0]-c\BK_{\frac{3}{4}}[z_0] &\equiv & 0~.
\label{defzero}
\eea
Next, using \refeq{defzero}, and recursion relations satisfied by the modified
Bessel functions \cite{abram}, we find, near $z_0$
\bea
\e^{-2x} & \approx & \left\{
\begin{array}{lr}
\frac{2}{3} z &\qquad  \mathrm{for}\quad c=0~,\\
z-z_0 &\qquad  \mathrm{for}\quad c>0~.
\end{array}
\right.
\eea
We also need to know the following asymptotic expansion near $z_0$
\bea
\BI_\frac{3}{4}[z]-c\BK_\frac{3}{4}[z] & \approx  &\left\{
\begin{array}{lr}
\frac{2^{3/4}}{3\Gamma (\frac{3}{4})}z^{3/4} &\qquad  \mathrm{for}\quad
c=0~,\\
\gamma  (z-z_0) &\qquad  \mathrm{for}\quad c>0~,
\end{array}
\right.
\label{expanI-K}
\eea
where we have defined the positive constant
$\gamma=\BI_{-\frac{1}{4}}[z_0]-c\BK_\frac{1}{4}[z_0]$, depending on $c$.
The IR behaviour of the solutions can now be written down, and
for simplicity we do so just for the $c>0$ solutions.
If we define $y=z-z_0$, we find (for $\psi\neq 0,\pi$) the IR limit
\bea
\dd s^2 &\approx & \dd\xi^2+\frac{2z_0}{g^2}\dd
\Omega^2_3+\frac{1}{g^2y}\left[\dd y^2+ y^2 (E_1^2+E_2^2)+\dd
\psi^2\right] \\
\vspace{3mm}
\e^{-2\Phi+2\Phi_0} &\approx& \gamma^2\sqrt{z_0} y\sin^2\psi~.
\eea
The form of the metric indicates that the singularity corresponds
to a linear distribution of fivebranes. This is parameterized by the angle
$\psi$ which varies in $0\leq\psi\leq\pi$ and it has therefore the topology
of a segment.
The origin of the $\sin^2\psi$ factor in the expression for the dilaton is
not clear, as one expects the same harmonic function, in this case $1/y$
appearing in the metric and $\e^{2\Phi}$. 
Presumably it means one needs to take the limit more
carefully. However, we can give some more evidence for the picture
advocated, by studying the limit near $\psi=0,\pi$ where
the above expansion is certainly not valid. 
In this case we find it useful to introduce a change of variables similar
to one performed in
\cite{gkmw}. Near $\psi=0$, for instance, let
\bea
y & = & \sqrt{\rho} \sin\frac{\alpha}{2}\nn
\psi & = & \sqrt{\rho} \cos\frac{\alpha}{2}
\eea
from which the expansions take the form
\begin{align}
\dd s^2 &\approx  \dd\xi^2+\frac{2z_0}{g^2}\dd
\Omega^2_3+\frac{1}{4g^2\rho^{3/2}\sin(\alpha/2)}\Big(\dd \rho^2+
\rho^2 \big[\dd\alpha^2 +\sin^2\alpha (E_1^2+E_2^2)\big]\Big) \\
\e^{-2\Phi+2\Phi_0} &\approx \gamma^2\sqrt{z_0}
\rho^{3/2}\sin(\alpha/2)
\end{align}
and note that $1/\rho^{3/2}\sin(\alpha/2)$ is harmonic in $\bbR^4$.
Keeping $\rho$ small, but finite, the above expressions show that the
singularity occurs for $\alpha=0$, which is the segment near its
end point.

\section{A probe computation}
\label{probecomput}

We would like to argue that the family of solutions
corresponding to good singularities ($c\geq 0$) should
describe a slice of the Coulomb branch of $D=3$ ${\cal N}=2$ pure
Yang--Mills theory. It is known that the perturbative
Coulomb branch of this theory is unstable due to the generation of
a superpotential \cite{ahw,deBoer,aha}. However, this superpotential
is induced by instanton effects,
and it has been observed in previous cases \cite{gkmw,bpp} that such effects
are not captured by the supergravity approximation. 
Thus, it is not unexpected that we can
find a supergravity solution dual to the perturbative
Coulomb branch of the gauge theory.

We shall provide evidence for our interpretation by probing
our solutions with a fivebrane. The $U(1)$ dynamics of the probe
fivebrane corresponds in the field theory language to
Higgsing the gauge group $SU(N)\to SU(N-1)\times U(1)$ and analysing
the dynamics of the $U(1)$ factor.
To properly treat the IR dynamics of the gauge theory on the NS fivebranes
one should switch to an S-dual description in terms of D-fivebranes
\cite{imsy}. The D5 brane solution is obtained via
the following transformation rules
\bea\label{dfivesol}
\Phi_\mathrm{D5} & = & -\Phi_\mathrm{NS5} ~, \nn
\dd s^2(\mathrm{D5}) & = & \e^{-\Phi_\mathrm{NS5}}
   \dd s^2(\mathrm{NS5}) ~, \nn
C_{(2)} & = & -B~ ,
\eea
where the quantities on the left hand side refer to the S-dual
dilaton, metric, and RR potential. Similarly, the corresponding
six-form potential dual to $C_{(2)}$ is given by $C_{(6)} = -\tilde
B$.

The effective action of a probe D5 brane in the string frame reads
\bea
\label{probeact}
S  & = & - \mu_5 \int \dd^6 y\; \e^{-\Phi_\mathrm{D5}}
\sqrt{-\det \left[ G_{ab}+ B_{ab} +
2\pi \alpha'{}F_{ab} \right]} \nonumber\\
&& \qquad + \mu_5\int  [ \exp (2\pi\alpha' {}F + B)
\vv \oplus_{n} C_{(n)} ]~,
\eea
where $\mu_5=(2\pi)^{-5}{\alpha'}^{-3}$, ${}F_{ab}$ is a world-volume
Abelian gauge field, $B$ is the NS two-form, $C_{(n)}$ are the RR
$n$-forms and it is understood that ten-dimensional fields are pulled
back to the six-dimensional world volume. The S-dual D5-brane solution has
$B=0$, and non-vanishing $C_{(2)}$ and $C_{(6)}$. However, one immediately
sees that the pullback of the three-form has no components on the three-sphere
on which the brane is wrapped, which means that there is no
Chern--Simons term in the corresponding gauge theory.

We choose the world-volume to have topology $\bbR^{1,2}\times S^3$ and
fix reparametrisation invariance by identifying the
world-volume coordinates with
$\{\xi^i,\tilde\psi,\tilde\theta,\tilde\phi\}$, where the
tilded angles are coordinates on the three-sphere on which the
brane is wrapped.
The four scalar fields $z$, $\psi$, $\theta$, and $\phi$
are then functions of these coordinates. The dynamics of the fivebrane
relevant for the three-dimensional gauge theory is obtained by letting
the fields depend only on the non-compact coordinates $\{\xi^i\}$.
In order to find which moduli do not break supersymmetry,
let us first consider the static part of the action, setting
\{$z$, $\psi$, $\theta$, $\phi$\} to constants and
${}F=0$. The contribution from the WZ part is easily obtained from
\refeq{NSsixform} and reads
\bea
S_\mathrm{WZ} & = & 2\pi^2\mu_5\left(\frac{2z}{g^2}\right)^{3/2}\int
\dd^3\xi\e^{-2\Phi}~.
\eea
The static contribution from the DBI determinant is in general rather
complicated due to contributions from the twisted part of the metric (contained
in $E_1$ and $E_2$), however it is not difficult to verify that it
cancels against the WZ part if we restrict to $\psi =0,\pi$ or
to $z=z_0$, in which case
the two terms vanish separately due to the vanishing of the dilaton at the
singularity. We therefore identify two distinct loci for supersymmetric motion
of the wrapped brane (cf. \cite{gkmw}).

We next consider non-zero velocities for the fields in order to find
the moduli space metric on the two loci. {}First, let us treat the 
gauge field. We are interested in
turning on a gauge field in the unwrapped directions of the brane.
After integrating over the three-sphere that the brane wraps, we can
dualise the $D=3$ $U(1)$ gauge-field (see e.g.\cite{schmi})
to get a compact scalar field. 
The $\phi$ and $\theta$ kinetic 
terms vanish upon restricting to the supersymmetric loci, 
leaving the following non trivial contributions in each of the two cases
\bea
S_\mathrm{I} & = & -\int \dd^3\xi \left[c_1 z^2
\Big(\BI_\frac{3}{4}[z]-c\BK_{\frac{3}{4}}[z]\Big)^2(\de z)^2
+c_2 z^{-3/2}(\de\sigma)^2 \right]\quad (\psi=0,\pi)~,
\eea
valid for $c\ge 0$, 
and
\bea
S_\mathrm{II} & = & -\int \dd^3\xi \bigg[c_1\gamma^2 z_0^2 \sin^2\psi(\de \psi)^2
+c_2 z_0^{-3/2}(\de\sigma)^2 \bigg]\quad (z=z_0)~,
\eea
valid for $c>0$. The constants $c_{1,2}$ are defined as follows
\bea
c_1 = \frac{\e^{-2\Phi_0}}{\pi^3 2^{7/2} g^5}~,
\quad\qquad c_2= 2^{3/2}\pi^3
g^3~.
\eea
Let us describe the resulting moduli space, focusing on the $c>0$ case.
Locus II has the topology of a cylinder and it covers the space-time
singularity. The moduli space metric on the cylinder is flat as can be
seen by employing the change of coordinates $\zeta=\cos\psi$, giving 
\bea
\dd s^2_{{\cal M}_{II}} & = & c_1 z_0^2 \gamma^2 \dd \zeta^2 + c_2 z_0^{-3/2}
\dd \sigma^2~.
\eea
On the other hand, locus I consists of two disconnected parts
($\psi=0$ and $\psi=\pi$), each of which is a cylinder whose
radius is finite near the singularity and goes to zero as $z\to \infty$.
By performing the change of coordinates
$\eta=1/2(z-z_0)^2$ we can use the expansion \refeq{expanI-K} to obtain the
following expression for the metric on locus I near the singularity
\bea
\dd s^2_{{\cal M}_{I}} \simeq c_1 z_0^2 \gamma^2 \dd \eta^2 + c_2 z_0^{-3/2}
\dd \sigma^2 \qquad\mathrm{for}\qquad z\to z_0~.
\eea
We notice that, in this limit, the radius of each of the two cylinders
of Locus I is identical to that of Locus II. 
This suggests, as in \cite{gkmw}, that we identify the three pieces
along the boundaries, resulting in a single
infinite ``cylinder'' both ends of which are shrinking to 
zero as $z\to\infty$. Note that the moduli space is smooth even where 
the space-time is singular.

\section{Geometry of wrapped fivebranes and generalised calibrations}
\label{gencal}

In this section we will further elucidate the types of geometry
that arise when IIB fivebranes wrap supersymmetric cycles in special
holonomy manifolds $X$. The results extend simply to type IIA by T-duality
and to type I in an obvious way. To be explicit one starts by
considering a probe brane in a geometry $\bbR^{1,p}\times (X\times
\bbR^q)$. Here the first factor represents the unwrapped brane
directions. The brane is taken to be wrapped on some cycle $\Sigma$ in
$X$. The factor $\bbR^q$ simply represents the remaining ``overall
transverse'' directions to the brane, making the total space
ten-dimensional. 

In all cases, in the string frame, the back-reaction of the brane on the
geometry preseves the flat $\bbR^{1,p}$ factor but promotes
$X\times\bbR^q$ to a non-trivial Riemannian manifold $Y$ with
non-vanishing dilaton and NS three-form. Moreover, in general, $Y$ has
two connections with totally anti-symmetric torsion given by
$\nabla^\pm=\nabla\pm \frac{1}{2}H$, as in \eqref{eq:gencon}. We will
see that in some cases both of these have special holonomy while in
others just one of them has special holonomy. The supersymmetry
transformations in type IIB are parametrised by a doublet of
Majorana-Weyl $SO(1,9)$ spinors $\bep=(\epsilon^-,\epsilon^+)$ with
the same chirality. In each case
the structure on $Y$ can be inferred from the projections on the
Killing spinors coming from, first, the special holonomy on
$X$ and, second, the additional projection implied by the presence of a
brane. The latter is different for $\epsilon^-$ and $\epsilon^+$. 
The cases where only one of $\nabla^-$ or $\nabla^+$ have special
holonomy on $Y$ occurs when the two different projections on 
$\epsilon^-$ and $\epsilon^+$ implied by the brane combined with 
the geometrical projections implied by $X$ only preserve 
$\epsilon^-$ or $\epsilon^+$ spinors. Cases where both $\nabla^+$ and 
$\nabla^-$ have special holonomy on $Y$ occur when
the combined projections preserve both $\epsilon^-$ and $\epsilon^+$ spinors.
In particular, if there
is an overall transverse direction when a fivebrane wraps the
supersymmetric cycle, i.e. $q$ is non-zero, then there are always
two structures.  

Let us first dicuss the seven-dimensional case where $Y$ has $G_2$
holonomy and then turn to the other cases. In this paper we have
presented a solution describing a fivebrane wrapping a SLAG
three-cycle in a Calabi--Yau threefold $X$. This is a case where there
is one overall transverse direction when a probe fivebrane wraps the
cycle. In the full solution, including the back-reaction, the
space no longer has a product structure, but the decomposition into
$X$ and an overall transverse direction is nonetheless still encoded in
the supersymmetry projections. In particular, the supersymmetry
projections for the parallel spinors of a
Calabi--Yau threefold can be written, as in
eqn.~\eqref{projection} with some relabelling, as 
\bea\label{CYproj}
\Gamma^{1234} \bep & = &\bep \nn
\Gamma^{3456} \bep & = & \bep~. 
\eea
If we decompose each $SO(1,9)$ Majorana-Weyl spinor $\epsilon^\pm$
into $SO(1,3)\times SU(3)\subset SO(3,1)\times Spin(6)$
representations, these projections imply that the preserved spinors
are $SU(3)$ singlets. As usual this gives eight supercharges or ${\cal
  N}=2$ in $D=4$. The presence of a probe fivebrane implies the extra
projection 
\bea
\Gamma^{1357} \bep & = & \tau_3\bep~.
\eea
As we have seen these projections are satisfied by the Killing spinors of
our particular solution. However, the analysis is quite general. 
Any supergravity solution representing a fivebrane wrapping a SLAG cycle 
will incorporate the same projections. Many features of the 
particular solution then go
through. First decompose each $SO(1,9)$ Majorana-Weyl spinor 
into $SO(1,2)\times Spin(7)$ via $\rep{16}_+\to (\rep{2},\rep{8})$, and
with a slight abuse of notation, let $\bep$ also denote the doublet
$(\epsilon^-,\epsilon^+)$ of seven-dimensional spinors. 
The above projections imply that for each $\epsilon^\pm$ there
is a $G_2$ subgroup of $Spin(7)$ with ${\bf 8}\to {\bf 7} +{\bf 1}$
with $\epsilon^\pm$ transforming as the singlet. Note that
the presence of $\tau_3$ implies that the projections are
different for each $\epsilon^\pm$ (related by simply reversing the sign
of the overall transverse tangent direction $e^7$) 
and hence the $G_2$ subgroups are different (related
by an outer automorphism). Since the Killing spinors satisfy 
$\nabla^-\epsilon^-=0$ and $\nabla^+\epsilon^+=0$ (see \refeq{gravitino}),
the holonomy of the two connections $\nabla^\pm$ are each separately
in $G_2$.  This implies that one can construct two different associative 
three-forms on $Y$, given by 
\begin{equation}
   \label{eq:phis}
   \phi^\pm_{\alpha\beta\gamma} = {\epsilon^\pm}^\trsp
        \gamma_{\alpha\beta\gamma} \epsilon^\pm~,
\end{equation}
where $\gamma^\alpha$ are seven-dimensional gamma matrices, and we
normalise $\epsilon^\pm$ by ${\epsilon^\pm}^\trsp\epsilon^\pm=1$. By
construction each associative threeform is covariantly constant 
but with respect to different connections with torsion
\begin{equation}
   \nabla^- \phi^- = 0 ~, \qquad
   \nabla^+ \phi^+ = 0~.    
\end{equation}
In summary, we see that the supersymmetry projections on the
spinors on the Calabi--Yau threefold $X$ together with the 
projection for the probe
fivebrane imply that $Y$ admits two different connections 
with $G_2$ holonomy and we have four conserved supercharges,
corresponding to ${\cal N}=2$ in $D=3$ on the unwrapped part
of the fivebrane world-volume. 

By contrast let us consider probe fivebranes wrapping associative
three-cycles in a seven-dimensional manifold $X$ of $G_2$
holonomy. Explicit solutions for this case were given
in~\cite{agk,Schvellinger:2001ib,Maldacena:2001pb}. In this case $q=0$
and there is no overall transverse direction. The $G_2$ holonomy of
$X$ implies the supersymmetry projections
\bea\label{XG2}
\Gamma^{1256} \bep& = & -\bep \nn
\Gamma^{1357} \bep& = & \bep\nn
\Gamma^{1467} \bep& = & \bep~.
\eea
As above, decomposing $SO(1,9)$ spinors into $SO(2,1)\times Spin(7)$,
these projections pick out the singlet representation of $G_2\subset Spin(7)$,
and give two conserved supercharges for $\epsilon^-$ and for $\epsilon^+$, 
corresponding to ${\cal N}=2$ in $D=3$. As before, 
the presence of a probe fivebrane implies the extra projection 
\bea
\Gamma^{1357} \bep & = & \tau_3\bep~.
\eea
Unlike the previous case, this is only compatible with the $G_2$
projections~\eqref{XG2} for $\epsilon^-$ and not for
$\epsilon^+$. Thus only two supercharges survive corresponding
to ${\cal N}=1$ in $D=3$ on the unwrapped part of the fivebrane. 
As a result, solutions that include the back-reaction on the
geometry give a new seven manifold $Y$ with torsion, and only one Killing
spinor $\nabla^-\epsilon^-=0$. That is, in contrast to the previous
case, there is now only one connection on $Y$ with $G_2$
holonomy. Consequently there is only one associative three-form $\phi^-$
satisfying, by construction, $\nabla^-\phi^-=0$. 
Indeed the explicit solutions constructed
in~\cite{agk,Schvellinger:2001ib,Maldacena:2001pb} are of this type and
it is worth emphasizing that the solutions in 
\cite{Schvellinger:2001ib,Maldacena:2001pb} are non-singular examples
of this kind of geometry.

Before considering the cases of other wrapped NS fivebranes, let us
return to the issue of generalised calibrations mentioned at the end of
section~\ref{susy}. Let us first recall the notion of calibration. On
a manifold of $G_2$ holonomy the associative three-form $\phi$ is a
calibration in that it satisfies the following conditions. First, if
pulled back onto any tangent three-plane it is less than or equal to
the induced volume-form evaluated on the three-plane. Secondly, 
it is closed $\dd\phi=0$. As a result a calibrated three-manifold
whose volume-form is equal to the pull-back of the associative
three-form is  minimal and, essentially because the associative
three-form has a spinorial construction, is supersymmetric. 

If the background also includes a form-field flux, the notion of
calibration must be generalised~\cite{gp,gpt} in order for it
to characterise supersymmetric cycles. In particular, $\phi$
is no longer closed, but rather $\dd\phi$ is related to the
flux. Recall that, in section~\ref{susy}, we showed that in our
solution $\phi$ was related to the dual NS potential $\tilde{B}$
together with the dilaton. The result~\eqref{eq:gencalG2}, can be
rewritten as  
\bea
\label{G2gencal}
H=\mp \e^{2\Phi}*_7 \dd(\e^{-2\Phi}\phi^\pm)
\eea
where $*_7$ is the Hodge dual operator on $Y$ and this is the correct
generalisation of the calibration condition $\dd\phi=0$ 
in the presence of both NS flux and a non-trivial dilaton. To see this,
recall that the action for a NS fivebrane with zero world-volume gauge field
strength and no background RR fields is given by 
\begin{equation}
   \label{eq:NSaction}
   S = - \mu_5 \int \dd^6y\;\e^{-2\Phi}\sqrt{-G} - \mu_5 \int \tilde{B}~,
\end{equation}
where $G$ is the induced world-volume metric.
Let us further consider the NS fivebrane probe to be static in
a spacetime of the form $\bbR^{1,2}\times Y$. One can then repeat the
same steps as in \cite{gp,gpt} to conclude that 
if the fivebrane has world-volume with volume form given by the pull-back 
of $\mathrm{Vol_3} \vv \e^{-2\Phi} \phi$ then, given the 
condition~\eqref{G2gencal}, it minimises the fivebrane static energy and, 
hence, $\phi$ is a generalized calibration and the probe configuration 
is supersymmetric. 

To see the connection with supersymmetry, we now show that
this condition can be derived in general from the
supersymmetry conditions. Reducing to seven dimensions,
the dilatino variation can be rewritten as 
\begin{equation}
   \delta\lambda = \partial_\alpha\Phi \gamma^\alpha\tau_3 \bep
        - \frac{1}{12}H_{\alpha\beta\gamma} 
             \gamma^{\alpha\beta\gamma}\bep = 0~.
\end{equation}
Given the symmetry properties of $\gamma^\alpha$, this implies that
\begin{equation}
\begin{aligned}
   \label{eq:commutator}
   \partial_\alpha\Phi 
          {\epsilon^-}^\trsp[A,\gamma^\alpha]_\mp\epsilon^-
      - \frac{1}{12}H_{\alpha\beta\gamma}
          {\epsilon^-}^\trsp[A,\gamma^{\alpha\beta\gamma}]_\pm\epsilon^-
      &= 0 \\
   \partial_\alpha\Phi 
          {\epsilon^+}^\trsp[A,\gamma^\alpha]_\mp\epsilon^+
      + \frac{1}{12}H_{\alpha\beta\gamma}
          {\epsilon^+}^\trsp[A,\gamma^{\alpha\beta\gamma}]_\pm\epsilon^+
      &= 0 \\
\end{aligned}
\end{equation}
where $A$ is an operator built out of gamma matrices and
$[\,\cdot\,,\,\cdot\,]_\pm$ 
refer to the anticommutator and commutator. Consider the case where 
$A=\gamma_{\alpha_1\alpha_2\alpha_3\alpha_4}$ and choose the upper
sign. Using the fact that in general either $\nabla^-\epsilon^-=0$ and
$\nabla^+\epsilon^+=0$ or only one vanishes, and with the orientation
on $Y$ given by $\epsilon^{\alpha_1\dots\alpha_7}=
{\epsilon^\pm}^\trsp\gamma^{\alpha_1\dots\alpha_7}\epsilon^\pm$, 
it is then easy to show that~\eqref{eq:commutator} implies the
generalized calibration condition~\eqref{G2gencal} for one or both of
the associated three-forms. In particular, when both $\phi^\pm$ exist,
each separately are generalized calibrations, while when there is just
a single $G_2$ structure, we have a unique generalized calibration. 

Thus far our discussion has focussed on the associative three-forms
$\phi^\pm$ on $Y$. However, one can always also construct the
corresponding co-associative four-forms
$*_7\phi^\pm_{\alpha_1\dots\alpha_4}=  
\epsilon^\pm\gamma_{\alpha_1\dots\alpha_4}\epsilon^\pm$ which are just
the Hodge duals of $\phi^\pm$. It is 
natural to ask what generlised calibration condition the 
co-associative form satisfies when the torsion and dilaton are non-vanishing. 
It is easy to show, 
taking $A=\gamma_{\alpha_1\dots\alpha_5}$ and the lower signs
in~\eqref{eq:commutator}, that 
\begin{equation}
   \label{eq:coass}
   \dd \left( \e^{-2\Phi} *_7\phi^\pm \right) = 0~.
\end{equation}
Note that, unlike the case of the associative form, the generalisation
does not involve the three-form $H$ and only involves the dilaton factor, 
which is necessary because of the corresponding factor in
the static energy as in~\eqref{eq:NSaction}. The general theory of
a single $G_2$  connection with totally anti-symmetric torsion has been 
recently developed in \cite{stefan}. A general expression 
for $d *_7\phi$ is given in \cite{stefan} that is consistent with \p{eq:coass}.
In addition our expression for the torsion \p{G2gencal} 
is consistent with the revised results of \cite{stefan}.

Let us discuss more briefly the geometries arising 
for other cases of wrapped NS fivebranes. 
First consider six-dimensional manifolds $Y$ with
connections with torsion of $SU(3)$ holonomy. The first way
in which these geometries arise is when 
fivebranes wrap two-cycles in a Calabi--Yau two-fold $X$.
In this case $q=2$ so there are two overall transverse directions
and when the back-reaction of the fivebrane on the geometry
is included the dimension of the non-trivial geometry jumps from
four to six. The projections on the spinors for $X$ imply
\begin{equation}
   \label{eq:CY2}
   \Gamma^{1234}\bep = \bep
\end{equation}
so that the preserved supersymmetries are singlets under the
$SU(2)$ holonomy on $X$. The projection due to the brane gives
\begin{equation}
   \label{eq:CY2b}
   \Gamma^{3456}\bep = \tau_3 \bep~.
\end{equation}
Comparing with~\eqref{CYproj} we see that these do indeed correspond
to the projections on parallel spinors of a manifold with $SU(3)$ holonomy, 
though clearly
with different $SU(3)$ structures for $\epsilon^+$ and $\epsilon^-$,
related by exchanging $e^5$ and $e^6$. Thus we conclude 
that on $Y$ the two connections $\nabla^\pm$ each have $SU(3)$
holonomy and that we can construct two commuting complex structures
$J^\pm$ with $\nabla^- J^-=0$ and $\nabla^+ J^+=0$. (Note that it can
be shown~\cite{strom}, in general, that for all the cases of $SU(N)$
holonomy with torsion that we discuss, the objects referred to as
complex structures constructed from the spinors
do indeed satisfy the conditions for $Y$ to be a complex manifold.) 
Since both $\epsilon^-$ and $\epsilon^+$ are each
separately conserved, the theory preserves $\mathcal{N}=2$ supersymmetry in
four dimensions. This kind of geometry was first noticed in a sigma
model context in \cite{ghr}. In addition we can repeat the generalised
calibration argument, taking $A=\gamma_{\alpha_1\alpha_2\alpha_3}$ in
the six-dimensional expression analogous to \eqref{eq:commutator}. We
find  
\begin{equation}
   \label{comp}
   H = \mp\e^{2\Phi}*_6\dd\left(\e^{-2\Phi} J^\pm\right)~,
\end{equation}
In fact, both of these structures were found in \cite{gkmw} for the particular
singular solutions presented in \cite{gkmw,zaf}. 

Six-dimensional geometries also arise when fivebranes wrap two-cycles
in Calabi--Yau threefolds, in which case $q=0$ and there are no
overall transverse directions. In this case, the spinor projections
from the geometry of $X$ were given in~\eqref{CYproj} above. However,
now the brane projection is given by~\eqref{eq:CY2b}. 
Clearly this is only consistent with the $X$ projections for
$\epsilon^-$, in which case the projections continue to imply $SU(3)$
holonomy. Thus we conclude that the geometry of $Y$ which includes the
back-reaction only has a single connection $\nabla^-$ with $SU(3)$
holonomy, and thus a single complex structure with $\nabla^-J^-=0$.
The preserved supersymmetry is now consequently $\mathcal{N}=1$ in $D=4$. 
The expression for the generalised calibration condition~\refeq{comp} is again
satisfied for $J^-$. Solutions of this type were constructed in
\cite{malnuntwo} and their geometry was discussed in \cite{papatse}.
Finally, it is worth noting that a different, but consistent,
expression for the torsion was found in \cite{ghr,strom,hull} 
(in the case of a single complex structure). The virtue of
\refeq{comp} is that this form generalises to cases
of $G_2$ or $Spin(7)$ holonomy. It is also worth noting that, as shown
in~\cite{strom,hull}, these geometries have holomorphic
three-forms of the form $\e^{-2\Phi}\omega^\pm$ where $\omega^\pm$ are
constructed from the Killing spinors. These are the generalised
calibrations corresponding to the special Lagrangian calibrations. 
These forms are closed, so just as for the co-associative forms
in the $G_2$ case above, the generalisation only involves the
dilaton.

Let us now turn to manifolds $Y$ with $SU(4)$ holonomy. These can be
obtained in four ways: from fivebranes wrapping (i) K\"{a}hler four-cycles in
Calabi--Yau threefolds, (ii) a product of two K\"{a}hler two-cycles in
a product of two Calabi--Yau two-folds, (iii) 
K\"{a}hler four-cycles in Calabi--Yau four-folds or (iv) complex
Lagrangian four-cycles in eight dimensional hyper-K\"ahler manifolds. 
The first and the third case are completely 
analogous to the cases with $SU(3)$ holonomy just described and it will
be convenient to discuss the fourth case later. 
Let us begin with the first case of four-cycles in threefolds. In this
case there are two overall transverse directions and
the projections on the parallel spinors
of $X$ are given
by~\eqref{CYproj} while the projection due to the brane gives
\begin{equation}
   \label{eq:CY4b}
   \Gamma^{5678}\bep = \tau_3\bep~.
\end{equation}
These are precisely the projections for $SU(4)$ holonomy on $Y$ though
with different structures for the two spinors $\epsilon^-$ and
$\epsilon^+$. That is we have two different complex structures $J^\pm$
satisfying $\nabla^-J^-=0$ and $\nabla^+J^+=0$. Again, the two
complex structures are related by exchanging the two overall
transverse directions $e^7$ and $e^8$. Note that this implies that the
eight-dimensional chirality of $\epsilon^-$ and $\epsilon^+$ are
different. As a result the theory preserves $(2,2)$ supersymmetry 
in the $D=2$ unwrapped world-volume dimensions. Again an expression 
for the generalised calibration can be
derived using $A=\gamma_{\alpha_1\alpha_2\alpha_3\alpha_4\alpha_5}$
in the eight-dimensional analog of~\eqref{eq:commutator} and reads    
\bea
\label{eq:calCY4}
   H = \frac{1}{2} \e^{2\Phi}*_8
      \dd\left(\e^{-2\Phi} J^\pm \wedge J^\pm\right)~. 
\eea
Note that the difference in chirality of $\epsilon^+$ and $\epsilon^-$
implies that $\epsilon^{\alpha_1\dots\alpha_8}=\epsilon^-\gamma^{\alpha_1\dots\alpha_8}\epsilon^-=-\epsilon^+\gamma^{\alpha_1\dots\alpha_8}\epsilon^+$ so there is no
overall sign factor in~\eqref{eq:calCY4}. 
The second case involves fivebranes wrapping a product of two
K\"ahler two-cycles in a product $X$ of two Calabi--Yau two-folds.
The projections for $X$ are
\begin{equation}
\label{eq:CY22}
\begin{aligned}
   \Gamma^{1234}\bep &= \bep \\
   \Gamma^{5678}\bep &= \bep~,
\end{aligned}
\end{equation}
while the presence of the brane wrapping a four-cycle tangent to the $1256$
directions implies 
\begin{equation}
   \label{eq:CY22b}
   \Gamma^{3478}\bep = \tau_3 \bep~.
\end{equation}
These imply we again have two $SU(4)$ structures. Note that in contrast
to the previous case, the preserved spinors have the same $SO(8)$ chirality,
\textit{i.e.} $\Gamma^{12345678}\bep = \bep$~, and this leads to these
configurations preserving $(4,0)$ supersymmetry.
Turning to the third case of fivebranes wrapping a K\"{a}hler four-cycle in a
Calabi--Yau four-fold $X$, the projections for $X$ now give
\begin{equation}
\label{eq:CY4}
\begin{aligned}
   \Gamma^{1234}\bep &= \bep \\
   \Gamma^{3456}\bep &= \bep \\
   \Gamma^{5678}\bep &= \bep~,
\end{aligned}
\end{equation}
while the presence of the brane implies the same
condition~\eqref{eq:CY4b} as above. This projection is only
compatible with the $X$ projections for $\epsilon^-$. Thus $Y$ has
a single $SU(4)$ complex structure $J^-$. Consequently only one chirality of
the spinor is preserved and the theory preserves $(2,0)$ supersymmetry in
$D=2$. Again the generalised calibration condition~\eqref{eq:calCY4}
is satisfied for $J^-$. As in the $SU(3)$ case, these geometries with
$SU(4)$ holonomy also have one or two holomorphic four-forms
$\e^{-2\Phi}\omega^\pm$, which are generalised calibrations
corresponding to what are special Lagrangian calibrations in case of
pure geometry where $H=\Phi=0$. 

Eight manifolds $Y$ with $Spin(7)$ holonomy can arise in
three ways. First, consider fivebranes wrapping co-associative
four-cycles in $X$ with $G_2$ holonomy, a case with one overall
transverse direction. The projections for $X$ were given
in~\eqref{XG2}. The brane projection reads
\begin{equation}
   \Gamma^{5678}\bep = \tau_3\bep~.
\end{equation}
The combined projections are then equivalent to two different
$Spin(7)$ structures on $Y$. In particular they can be rewritten as 
\begin{equation}
   \label{spin7}
   \begin{aligned}
   \Gamma^{1234}\bep &= \bep \\
   \Gamma^{3456}\bep &= \bep \\
   \Gamma^{5678}\bep &= \tau_3\bep \\
   \Gamma^{1357}\bep &= \bep~.
   \end{aligned}
\end{equation}
Again the two structures are related by reversing the overall
transverse tangent direction $e^8$ and give two different Cayley structures
$\Omega^-$ and $\Omega^+$ on $Y$ with $\nabla^-\Omega^-=0$ and
$\nabla^+\Omega^+=0$. Note in particular that the eight-dimensional
chirality of $\epsilon^-$ and $\epsilon^+$ on $Y$ are different. 
This means that 
the theory preserves $(1,1)$ supersymmetry in $D=2$. As before
one can derive a generalized calibration condition by taking
$A=\gamma_{\alpha_1\alpha_2\alpha_3\alpha_4\alpha_5}$ in the analog
of~\eqref{eq:commutator}, giving 
\bea
\label{eq:cal71}
   H = \e^{2\Phi}*_8 \dd\left(\e^{-2\Phi}\Omega^\pm\right)~. 
\eea
As above the difference in chirality of $\epsilon^-$ and $\epsilon^+$
means there is no overall sign factor. 

The second way to obtain $Y$ with $Spin(7)$ holonomy is by wrapping
SLAG four-cycles in a Calabi--Yau fourfold $X$ with $SU(4)$
holonomy. The projections for $X$ are given by~\eqref{eq:CY4}. The
brane projection reads
\begin{equation}
   \Gamma^{1357}\bep = \tau_3\bep~.
\end{equation}
Again this leads to two different $Spin(7)$ structures $\Omega^\pm$ on
$Y$. However, in contrast to the last case, these structures have the
\textit{same} eight-dimensional chirality. Thus this theory preserves
$(2,0)$ supersymmetry in $D=2$. Note that this is also
the first case of $Y$ with a pair of structures when there were no
overall transverse directions. As in the previous case, both
structures satisfy the generalised calibration
condition~\eqref{eq:cal71}, except since $\epsilon^-$ and $\epsilon^+$
have the same chirality there is an overall sign factor
\bea
\label{eq:cal72}
   H = \mp\e^{2\Phi}*_8 \dd\left(\e^{-2\Phi}\Omega^\pm\right)~. 
\eea

The last case with $Spin(7)$ holonomy comes from wrapping Cayley
cycles in $X$ with $Spin(7)$ holonomy. The projections from $X$ are
\begin{equation}
   \begin{aligned}
   \Gamma^{1234}\bep &= \bep \\
   \Gamma^{3456}\bep &= \bep \\
   \Gamma^{5678}\bep &= \bep \\
   \Gamma^{1357}\bep &= \bep~,
   \end{aligned}
\end{equation}
and the brane gives
\begin{equation}
   \Gamma^{5678}\bep = \tau_3\bep~.
\end{equation}
This is only consistent with the other projections for
$\epsilon^-$. Thus there is only one Cayley structure
$\Omega^-$ on $Y$, again satisfying the generalised calibration
condition~\eqref{eq:cal71}. Since only one
chirality of spinor survives, the theory preserves $(1,0)$
supersymmetry in $D=2$. 

Finally, let us discuss fivebranes wrapping complex
Lagrangian four-cycles in hyper-K\"ahler manifolds. These are
four-cycles that are complex with respect to one complex
structure and special Lagrangian with respect to the other two.
Solutions for M-fivebranes wrapping these cycles were recently given in
\cite{Gauntlett:2001jj}, and we will use the analysis at the beginning
of section~2 of that paper. Let the projections corresponding to the 
$Sp(2)$ singlets of a hyper-K\"ahler manifold $X$ be given by
\bea\label{elprojs}
\Gamma^{1256}\bep=\bep\nn
\Gamma^{3478}\bep=\bep\nn
(1-\Gamma^{1234}-\Gamma^{1458}-\Gamma^{2358})\bep&=0~.
\eea     
These preserve $(6,0)$ supersymmetry in $D=2$. One can easily
show~\cite{Gauntlett:2001jj} that the chirality of the three 
preserved $\epsilon^-$ spinors with respect 
to $\Gamma^{1234}$, $\Gamma^{1458}$ and $\Gamma^{2358}$ can be taken to be
$(+,+,-)$, $(+,-,+)$ and $(-,+,+)$, respectively, and similarly
for the three preserved $\epsilon^+$ spinors.
The four-cycle tangent to the $1234$ directions is complex Lagrangian 
and the corresponding projections on the spinors can be written
\bea
\Gamma^{1234}\bep=\tau_3\bep~.
\eea
We see that the configuration preserves 
the first two $\epsilon^-$ spinors and the third
$\epsilon^+$ spinor, giving $(3,0)$ supersymmetry in $D=2$.
The two $\epsilon^-$ spinors satisfy the same projections
as those preserved by a fivebrane wrapping a K\"ahler four-cycle in a
Calabi--Yau four-fold and imply that $\nabla^-$ has $SU(4)$ holonomy.
On the other hand, the single preserved $\epsilon^+$ spinor 
implies that $\nabla^+$ has $Spin(7)$ holonomy.

We have summarised the results of this section in the following table.
We have also included the amount of supersymmetry preserved in
type IIA and type I supergravity. For type IIB and type IIA the
amount of preserved supersymmetry is the same irrespective
of the orientation of the fivebrane. The cases where this is not true
for the type I theory are indicated in the last column of the table.
\begin{table}[!th]
\begin{center}
\bigskip
\footnotesize
\setlength{\tabcolsep}{0.45em}
\begin{tabular}{|cc|ccc|cccc|}
\hline
\multicolumn{2}{|c|}{Probe configuration}  &
\multicolumn{3}{|c|}{Geometry of $Y$} &
\multicolumn{4}{|c|}{World volume susy}\\ 
Cycle~$\Sigma$ & $X$ &  dim($Y$) & Hol($\nabla^{-}$) & Hol($\nabla^{+}$) &
$ D$ & ${\cal N}_\mathrm{IIB}$ & ${\cal N}_\mathrm{IIA}$ & ${\cal
  N}_\mathrm{I}$\\ 
\hline\hline
\Ka-2   & \CY$_2$  &      6         &  $SU(3)$     &    $SU(3)$ 
& 4 & 8 & 8 & 4\\
\hline
\Ka-2   & \CY$_3$  &      6         &  $SU(3)$     &    $SO(6)$ 
& 4 & 4 & 4 & 4 or none\\
\hline
SLAG-3 & \CY$_3$  &      7         &   $G_2$     &    $G_2$ 
& 3 & 4 & 4 & 2\\
\hline
Associative & \gman  &      7         &  $G_2$  &    $SO(7)$  
& 3 & 2 & 2 & 2 or none\\
\hline
\Ka-4   &  \CY$_3$ &      8         &  $SU(4)$  &  $SU(4)$  
& 2 & (2,2) & (4,0) & (2,0)\\
\hline
(\Ka-2)$^2$  &  \CY$_2\times$\CY$_2$ & 8 & $SU(4)$ & $SU(4)$ 
& 2 & (4,0) & (2,2) & (2,0)\\
\hline
$\mathbb{C}$LAG-4    & \HK  &      8  &  $SU(4)$ & \spin 
& 2 & (3,0) & (2,1) & (2,0) or (1,0)\\
\hline
\Ka-4   &  \CY$_4$ &      8         &  $SU(4)$  &  $SO(8)$  
& 2 & (2,0) & (2,0) & (2,0) or none\\
\hline
SLAG-4& \CY$_4$ &      8         &  \spin      & \spin 
& 2 & (2,0) & (1,1) & (1,0)\\
\hline
Co-associative &  \gman  &    8   &  \spin      &  \spin 
& 2 & (1,1) & (2,0) & (1,0)\\
\hline
Cayley  & \spin &      8         &  \spin      &  $SO(8)$ 
& 2 & (1,0) & (1,0) & (1,0) or none\\
\hline
\end{tabular}
\normalsize
\end{center}
\caption{Holonomy and supersymmetry of supergravity solutions for 
wrapped NS five-branes.}
\end{table}

Note that supergravity solutions describing intersecting branes calibrated by
quaternionic calibrations were discussed in~\cite{gphyper} where the
holonomy of the connections $\nabla^\pm$ was also deduced. Note that
unlike the cases we have been considering, these quaternionic cycles 
are necessarily linear~\cite{dadok}. 

\section{Discussion}

We have presented a $D=10$ 
supergravity solution describing IIB fivebranes wrapping
a SLAG three-cycle and discussed how it can be interpreted as a gravity dual
of pure ${\cal N}=2$ super-Yang-Mills theory in $D=3$. It would be very
interesting to extend this work to find gravity solutions dual to
${\cal N}=2$ theories with a Chern--Simons term since these can have
supersymmetric confining vacua. A Chern-Simons term arises 
when there is NS three-form flux on the SLAG three-sphere 
\cite{agk}. In order to construct such solutions within
gauged supergravity, one needs to switch on the $D=7$ three-form.
However, upon inspection of the equations of motion \cite{clp} 
it is not difficult 
to see that one has to go beyond the $SO(3)$ ansatz that we considered
in this paper.

We have also discussed some aspects of the geometry arising
when IIB fivebranes wrap supersymmetric cycles. In particular we argued
that in some cases one obtains geometries with both of the
connections with torsion, $\nabla^\pm$, having special holonomy 
and in others only one of them does. {}For each case we also
elucidated the appropriate notion of generalised calibration. 
Explicit supergravity solutions for several different cases of 
fivebranes wrapping supersymmetric cycles have now been found. 
It seems straightforward to find solutions for all cases, by
first constructing them in gauged supergravity, and these will
provide explicit examples of the remaining geometries that we discussed.
It would be interesting to see if the new cases have a dual 
field theory interpretation.

\section*{Acknowledgements}
We thank {}Fay Dowker, Jan Gutowski, Chris Hull,
and Stathis Pakis for helpful discussions.
All authors are supported in part by PPARC through SPG $\#$613.
JPG thanks EPSRC for partial support. DW is also supported by the
Royal Society.  
\appendix


\begin{thebibliography}{99}


\bibitem{bvs}
M.~Bershadsky, C.~Vafa and V.~Sadov,
``D-Branes and Topological {}Field Theories,''
Nucl.\ Phys.\  {\bf B463} (1996) 420,
hep-th/9511222.


\bibitem{malnun}
J.~Maldacena and C.~Nunez,
``Supergravity description of field theories on curved manifolds and a no  go
theorem,''
Int.\ J.\ Mod.\ Phys.\ A {\bf 16}, 822 (2001)
hep-th/0007018.

\bibitem{malnuntwo}
J.~M.~Maldacena and C.~Nunez,
``Towards the large $N$ limit of pure $\mathcal{N} = 1$ super Yang Mills,''
Phys.\ Rev.\ Lett.\  {\bf 86}, 588 (2001)
hep-th/0008001.

\bibitem{fs}
B.~Brinne, A.~{}Fayyazuddin, S.~Mukhopadhyay and D.~J.~Smith,
``Supergravity M5-branes wrapped on Riemann surfaces and their Q{}FT duals,''
JHEP {\bf 0012}, 013 (2000)
hep-th/0009047.

\bibitem{agk}
B.~S.~Acharya, J.~P.~Gauntlett and N.~Kim,
``{}Fivebranes wrapped on associative three-cycles,'', to appear in
Phys. Rev. {\bf D}
hep-th/0011190.


\bibitem{no}
H.~Nieder and Y.~Oz,
``Supergravity and D-branes wrapping special Lagrangian cycles,''
JHEP {\bf 0103}, 008 (2001)
hep-th/0011288.

\bibitem{gkw}
J.~P.~Gauntlett, N.~Kim and D.~Waldram,
``M-fivebranes wrapped on supersymmetric cycles,''
Phys.\ Rev.\ D {\bf 63}, 126001 (2001)
hep-th/0012195.

\bibitem{Nunez:2001pt}
C.~Nunez, I.~Y.~Park, M.~Schvellinger and T.~A.~Tran,
``Supergravity duals of gauge theories from $F(4)$ gauged supergravity in six
dimensions,''
JHEP {\bf 0104}, 025 (2001)
hep-th/0103080.

\bibitem{Edelstein:2001pu}
J.~D.~Edelstein and C.~Nunez,
``D6 branes and M-theory geometrical transitions from gauged  supergravity,''
JHEP {\bf 0104}, 028 (2001)
hep-th/0103167.


\bibitem{Schvellinger:2001ib}
M.~Schvellinger and T.~A.~Tran,
``Supergravity duals of gauge field theories from $SU(2)\times U(1)$ gauged
supergravity in five dimensions,''
JHEP {\bf 0106}, 025 (2001)
hep-th/0105019.


\bibitem{Maldacena:2001pb}
J.~Maldacena and H.~Nastase,
``The supergravity dual of a theory with dynamical supersymmetry  breaking,''
JHEP {\bf 0109}, 024 (2001)
hep-th/0105049.


\bibitem{gkpw}
J.~P.~Gauntlett, N.~Kim, S. Pakis and D.~Waldram,
``Membranes wrapped on holomorphic curves,'' to appear in Phys. Rev. {\bf D},
hep-th/0105250.


\bibitem{Hernandez:2001bh}
R.~Hernandez,
``Branes wrapped on coassociative cycles,''
hep-th/0106055.


\bibitem{gkmw}
J.~P.~Gauntlett, N.~Kim, D.~Martelli and D.~Waldram,
``Wrapped fivebranes and $\mathcal{N}=2$ super Yang-Mills theory,'' to
appear in Phys. Rev. {\bf D},
hep-th/0106117.


\bibitem{zaf}
{}F.~Bigazzi, A.~L.~Cotrone and A.~Zaffaroni,
``$\mathcal{N} = 2$ gauge theories from wrapped five-branes,''
hep-th/0106160.


\bibitem{Gomis:2001vg}
J.~Gomis and T.~Mateos,
``D6 branes wrapping Kaehler four-cycles,''
hep-th/0108080.


\bibitem{Gauntlett:2001jj}
J.~P.~Gauntlett and N.~Kim,
``M-fivebranes wrapped on supersymmetric cycles. II,''
hep-th/0109039.


\bibitem{ahw}
I.~Affleck, J.~A.~Harvey and E.~Witten,
``Instantons And (Super)Symmetry Breaking In (2+1)-Dimensions,''
Nucl.\ Phys.\ B {\bf 206}, 413 (1982).


\bibitem{deBoer}
J.~de Boer, K.~Hori and Y.~Oz,
``Dynamics of $\mathcal{N} = 2$ supersymmetric gauge theories in three
dimensions,'' Nucl.\ Phys.\ B {\bf 500}, 163 (1997)
hep-th/9703100.

\bibitem{aha}
O.~Aharony, A.~Hanany, K.~Intriligator, N.~Seiberg and M.~J.~Strassler,
``Aspects of $\mathcal{N} = 2$ supersymmetric gauge theories in three
dimensions,'' Nucl.\ Phys.\ B {\bf 499}, 67 (1997)
hep-th/9703110.

\bibitem{ohta}
K.~Ohta,
``Supersymmetric index and s-rule for type IIB branes,''
JHEP {\bf 9910}, 006 (1999)
hep-th/9908120.

\bibitem{chamvolk}
A.~H.~Chamseddine and M.~S.~Volkov,
``Non-Abelian vacua in $D = 5$, $\mathcal{N} = 4$ gauged supergravity,''
JHEP {\bf 0104}, 023 (2001)
hep-th/0101202.


\bibitem{papatse}
G.~Papadopoulos and A.~A.~Tseytlin,
``Complex geometry of conifolds and 5-brane wrapped on 2-sphere,''
Class.\ Quant.\ Grav.\  {\bf 18}, 1333 (2001)
hep-th/0012034.

\bibitem{gr}
J.~Gomis and J.~G.~Russo,
``$D=2+1$ $\mathcal{N}=2$ Yang-Mills Theory {}From Wrapped Branes,''
hep-th/0109177.

\bibitem{LST}
N.~Seiberg,
``New theories in six dimensions and matrix description of M-theory on  $T^5$
and $T^5/Z_2$,''
Phys.\ Lett.\ B {\bf 408}, 98 (1997)
hep-th/9705221.


\bibitem{maclean}
R.~C.~McLean,
``Deformations of calibrated submanifolds,''
Comm. Anal. Geom. {\bf 6} (1998) 705-747.

\bibitem{salsez}
A.~Salam and E.~Sezgin,
``$SO(4)$ Gauging Of $\mathcal{N}=2$ Supergravity In Seven-Dimensions,''
Phys.\ Lett.\ B {\bf 126}, 295 (1983).

\bibitem{clp} M. Cveti\u{c}, H. L\"u and C.N. Pope,
``Consistent Kaluza-Klein Sphere Reductions,''
Phys. Rev. {\bf D62} (2000) 064028,
hep-th/0003286.

\bibitem{abram}
M.~Abramowitz and I.~A.~Stegun (editors),
\textit{Handbook of mathematical functions},
Dover Publications, New York (1965).

\bibitem{bpp}
A.~Buchel, A.~W.~Peet and J.~Polchinski,
``Gauge dual and noncommutative extension of an $\mathcal{N} = 2$
supergravity solution,'' 
Phys.\ Rev.\ D {\bf 63}, 044009 (2001)
hep-th/0008076.

\bibitem{imsy}
N.~Itzhaki, J.~M.~Maldacena, J.~Sonnenschein and S.~Yankielowicz,
``Supergravity and the large $N$ limit of theories with sixteen supercharges,''
Phys.\ Rev.\ D {\bf 58} (1998) 046004
hep-th/9802042.

\bibitem{gp}
J.~Gutowski and G.~Papadopoulos,
``AdS calibrations,''
Phys.\ Lett.\ B {\bf 462} (1999) 81
hep-th/9902034.

\bibitem{gpt}
J.~Gutowski, G.~Papadopoulos and P.~K.~Townsend,
``Supersymmetry and generalized calibrations,''
Phys.\ Rev.\ D {\bf 60} (1999) 106006
hep-th/9905156.

\bibitem{schmi}
C.~Schmidhuber,
``D-brane actions,''
Nucl.\ Phys.\ B {\bf 467} (1996) 146
hep-th/9601003.

\bibitem{stefan}
T.~Friedrich and S.~Ivanov,
``Parallel spinors and connections with skew-symmetric torsion in string  theory,''
math.dg/0102142.

\bibitem{strom}
A.~Strominger,
``Superstrings With Torsion,''
Nucl.\ Phys.\ B {\bf 274} (1986) 253.

\bibitem{ghr}
S.~J.~Gates, C.~M.~Hull and M.~Rocek,
``Twisted Multiplets And New Supersymmetric Nonlinear Sigma Models,''
Nucl.\ Phys.\ B {\bf 248} (1984) 157.



\bibitem{hull}
C.~M.~Hull,
``Superstring Compactifications With Torsion And Space-Time Supersymmetry,''
In {\it  Turin 1985, Proceedings, Superunification and Extra Dimensions}, 
347-375.


\bibitem{gphyper}
G.~Papadopoulos,
``Brane solitons and hypercomplex structures,''
math.dg/0003024.

\bibitem{dadok}
J. Dadok, F.R. Harvey and F. Morgan,
``Calibrations on $R^8$,'' Trans. Am. Math. Soc.
{\bf 307} (1988) 1.   

\end{thebibliography}
\end{document}